\documentclass[12pt]{article}
\usepackage{graphicx}

\def\hn{{\it n}-heptane+nitrobenzene}
\def\hnform{${\rm C}_{7}{\rm H}_{16}+{\rm C}_{6}{\rm H}_{5}{\rm
N}{\rm O}_{2}$}

\def\op{order parameter}

\def\Tc{$T_{\rm C}$}

\long\def\symbolfootnote[#1]#2{\begingroup%
\def\thefootnote{\fnsymbol{footnote}}\footnote[#1]{#2}\endgroup}

\begin{document}


\begin{center}
    {\bf {\Large 
Coexistence curve of the \hn\ mixture near its consolute
point measured by an optical method}}
\end{center}
\vspace{5 mm}

\begin{center}
Nicola Fameli\symbolfootnote[1]{Corresponding author, presently at : 
Department of Anesthesiology,
Pharmacology and Therapeutics, The University of British
Columbia, 2176, Health Sciences Mall, Vancouver, B. C., Canada V6T
1Z3; Electronic address: {\tt fameli@physics.ubc.ca}}, David A. Balzarini\\{\it Department of Physics and Astronomy,} \\
{\it The University of British Columbia,}\\
{\it 6224 Agricultural Road, Vancouver, B. C., Canada V6T 1Z1}
\end{center}
\vspace{150mm}

\pagebreak

{\centerline{\Large Abstract}}
\vspace{5mm}

We have measured the coexistence curve of the binary liquid mixture
\hn\ (\hnform) near its consolute point using an optical method. In
particular, the critical exponent $\beta$ describing the coexistence
curve was measured for this system. Previous experimental values of
$\beta$ for \hn\ were higher than the typical theoretically
calculated value, an unusual, although not unique, occurrence. In an
effort to study this discrepancy, we have used an improved
experimental apparatus for our measurements. We have taken special
care to minimize temperature gradients and maximize the temperature
stability of our thermal control system. We have also exploited
features of a known optical method to analyze, thoroughly, sources of
systematic errors. We measured an apparent value of 
$\beta$ as $0.367\pm 0.006$ and by a
careful study of the known sources of error we find that they are not
able to remove the discrepancy between the measured and the
theoretical values of $\beta$. We also measured the critical
temperature of the system at $T_{\rm C}=(291.80\pm 0.02)$~K
(18.65~$^{\circ}$C).



\pagebreak

\section{Introduction}
One of the most salient features of critical phenomena is the
universal behaviour of the so-called critical
exponents. In
this work, we report results from an experiment, in which we measured
the critical exponent
$\beta$ describing the coexistence curve of the binary liquid
\hn. The former being lighter floats on the latter when the system is
below its critical consolute temperature of about 18.9~$^{\circ}$C~\cite{merck}. 
Theoretical studies predict the values of the
critical exponents to a very high degree of accuracy and several
experimental approaches can verify those
predictions (reviewed in \cite{kumar83, sengers86}). 
The critical phenomenon in question is the miscibility of the two
liquids as the temperature is varied through the critical, or 
consolute, temperature $T_{\rm C}$. Typically, binary fluids exist
as two immiscible phases below $T_{\rm C}$, while above the critical
temperature, the two liquids become completely miscible and only a
single liquid phase exists~\cite{invcoexnote}. If we call the two
phases $U$ and $L$ (for `upper' and `lower', respectively), the order
parameter of choice for
these systems is usually the difference between the concentration of
one of
the species (e.g., {\it n}-heptane) $\phi_{\rm H}$ in  phase $U$
and its concentration in the phase $L$~\cite{ordparamnote}:
$\Delta\phi_{\rm H}=\phi_{\rm H, U}-\phi_{\rm H, L}$.
Introducing the
reduced temperature $t=(T_{\rm C}-T)/T_{\rm C}$, the shape of the
coexistence curve very close to the critical point should be described
by the simple scaling equation:
$\Delta\phi=B_{0}t^{\beta}$,
while further
from the critical point correction-to-scaling terms are
needed:
$\Delta\phi=B_{0}t^{\beta}(1+B_{1}t^{\Delta}+B_{2}t^{2\Delta}+\cdots)$,
where
the amplitudes $B_{0}$, $B_{1}$, $B_{2}, \ldots\;$ are
system-dependent coefficients to be determined, and $\Delta$ is the
correction-to-scaling critical exponent~\cite{wegner72}. The
theoretical value of $\beta$ is predicted to be approximately $\beta=0.326\pm 
0.002$~\cite{guida98}. 
One way to test the validity of
theoretical predictions on the exponent $\beta$ is to measure the
coexistence curve of a binary liquid mixture by optical
interferometry, which is the method employed in this
work~\cite{db74}. 

The decision to measure the coexistence curve of a binary liquid
mixture stems from the need to clarify discrepancies between the
theoretical values of $\beta$ and experimental results first obtained
in the 1970s~\cite{stein73, kumar83}.
It appears quite clearly from earlier literature that until the late 
1970s or even early 1980s the typical accepted experimental values
for the critical exponent $\beta$ describing the order parameter 
dependence on temperature in binary mixtures was somewhat 
larger than the expected theoretical value for that
exponent~(\cite{stein73, kumar83, jay81, db79}). In a 
compilation of experimental data on binary liquid mixtures, Stein and 
Allen~\cite{stein73} found that the data of the systems they analyzed
were all consistent with the critical exponent $\beta=0.34\pm0.01$.
The trend of other experimental studies of coexistence
curves of binary liquid mixtures seemed instead to be confirming the
renormalization group theoretical result that the critical exponent 
$\beta$ should range between $0.325<\beta<0.327$ \cite{db74, greer76}. 

Following the relative uncertainty of the experimental results
obtained 
on binary liquids, coupled with the fact that one of the systems
displaying a greater than expected value of $\beta$ was \hn,
it seemed fit and interesting to perform a new study of the
coexistence curve of that system, with a new and improved
experimental apparatus. We
believe that the thermal control system employed for this work allows
greater temperature stability and uniformity. Moreover, the
calibration of the thermometers was carried out more accurately and
features of our optical technique were used to monitor more carefully
the status of
equilibrium (or lack thereof) of the system. All of
these aspects of the experimental procedure were treated with greater
care than in earlier studies of \hn, so as to obtain as
accurate as possible a value of the critical exponent $\beta$.  

\section{Experimental features}
\subsection{Optical observation of fluid critical phenomena}\label{optical}
\begin{figure}[tb]\centering
\includegraphics[scale=0.5]{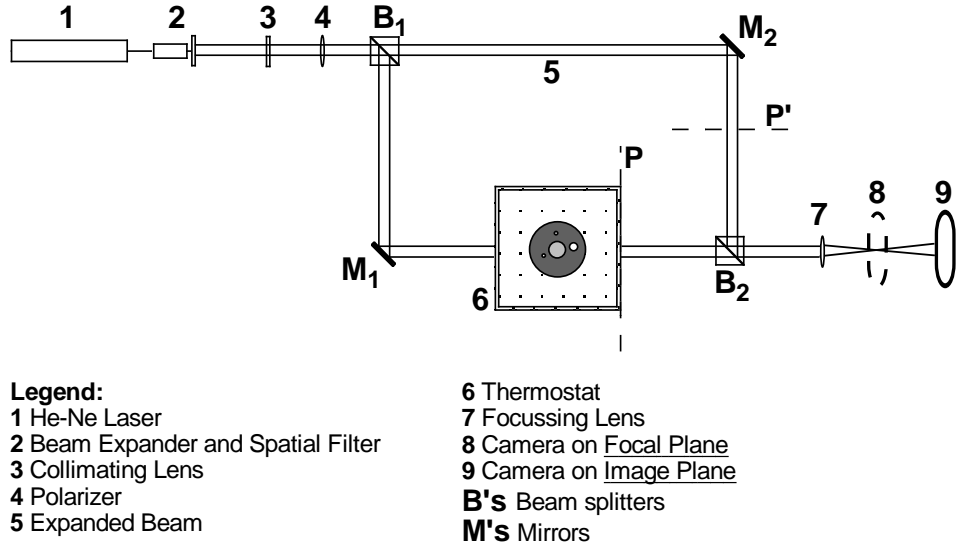}
\caption{Schematic diagram of the experimental optical setup used for 
the focal plane and image plane detection techniques. See figure 
legend for explanation of various parts of the setup.}\label{machzehnder}
\end{figure}

\begin{figure}[tb]\centering
\includegraphics[scale=0.5]{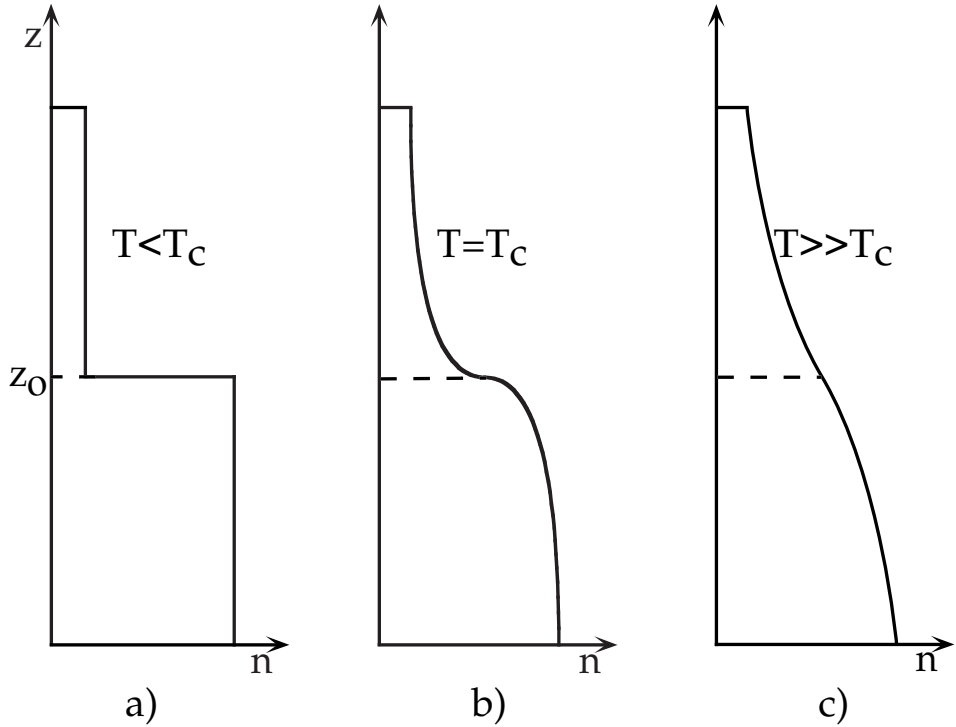}
\caption{Illustration of the behaviour of the refractive index of a
binary mixture as function of cell height at $T<T_{\rm C}$ (a), at
$T_{\rm C}$ (b), and  $T>>T_{\rm C}$. $z_{0}$ marks the position of
the meniscus between the two  phases, below $T_{\rm
C}$.}\label{nprofile}
\end{figure}

The experimental method is based on the detection of variations that
occur in the index of refraction of the liquid contained in a cell as
its temperature varies. Our measurements are based on two optical 
detection techniques: the focal plane and the image plane 
techniques (Fig.~\ref{machzehnder}). These techniques were already 
used for similar experiments in the past, and we refer to those 
studies for details~\cite{db74, dbthesis, db68}.
In the focal plane technique, we exploit the refractive index gradient 
occurring as the sample's temperature is raised from below 
to above the critical temperature \Tc. In the process, a fraction of
the heptane diffuses into the nitrobenzene and {\it vice versa}. As a
result, the refractive index profile, discontinuous below \Tc, takes
instead a sigmoidal shape (Fig.~\ref{nprofile}).
The refractive index profile of the sample is probed using a
coherent beam of light from a 632.8 nm He-Ne laser, with the beam
expanded to approximately 25-mm in diameter and collimated. Light
passing through the cell is bent due to the refractive index gradient
in the sample. Rays encountering the sigmoidal profile at points with
equal absolute value of the curvature are bent through equal angles.
If the light is then focussed on a screen by a lens, a diffraction
pattern appears in the focal plane of the lens. 
This diffraction pattern is recorded on photographic film. The
number of fringes, $N$, observed in the diffraction pattern 
at each temperature is directly proportional to the
difference in refractive index $\Delta n$ between the two phases:
$\Delta
n=n_{\rm U}-n_{\rm L}$. It can be shown that, to a good approximation
under
our experimental conditions~\cite{db74, jacobs76}, $\Delta n$ is
proportional to the order parameter of the binary fluid system: 
$\Delta n=k\Delta\phi$ ($k$ was previously measured as 6.045 
\cite{sclavo}; we will discuss this issue further in 
section~\ref{mixingvolume}). Therefore, a $\Delta\phi$-versus-temperature graph
represents the coexistence curve of the binary mixture and it is
suitable for an indirect measurement of the exponent $\beta$.

In the image plane technique, the plane of the recording photographic 
film is placed on the image plane of the lens also used 
in the focal plane technique, and the 
whole refractive index profile of the sample is imaged on the film. 
As illustrated in Fig.~\ref{machzehnder}, 
this is achieved by placing the thermostat and sample in one of the 
beams of a Mach-Zehnder interferometer,
while the other beam travels through air.
As 
the refractive index varies considerably with 
cell height, when the  temperature of the sample is in the 
neighbourhood of the critical temperature, the bottom part of the 
sample has a larger optical thickness, hence the light of the sample 
beam that traverses the bottom part of the sample is slowed down more 
than the light passing through the top part.
Therefore, when the waves at P and ${\rm P}'$ are recombined
and 
imaged by means of the lens, horizontal interference fringes are 
observed at the image plane of the lens. 
The index of refraction as a function of height is mapped in this
way. 
We used this technique to study the sample equilibration time 
at a set temperature.

\subsection{Sample preparation}\label{samples}
The liquids were purchased from Fisher Scientific 
with stated purities of 99\% for nitrobenzene and 95\% for {\it
n}-heptane. 
They were distilled neat (the former under reduced pressure) and
brought to an 
estimated purity of better than 99.9 \%, as estimated by gas
chromatography as 
well as by NMR. 
A mixture was prepared in the proportions of 49.6 wt\% {\it
n}-heptane and 
50.4 wt\% nitrobenzene, as suggested by earlier 
studies~\cite{brumb65} and the sample was prepared
following a  
method devised and used by this laboratory in the past, with very
good 
results~\cite{db74}. 

Several different sample cells were prepared. 
The cells have a vertically elongated 
parallelepipedal shape, about 5 cm in height, 1 cm in width, and 
different depths (the dimension parallel to the He-Ne beam), 
depending on the light paths wanted for the 
measurements. In order to observe both the region 
far from the critical point 
and the near-critical region with relative ease, the  sample 
cells were chosen to have light 
paths of 1, 2, 5, and 10~mm (the tolerance of the light 
paths as quoted by the manufacturer is 0.01~mm; the heights of the 
fluid in the 1-, 2-, 5-, and 10-mm samples were 34, 35, 37, and 38.5 
mm, respectively).
In our focal plane measurements, 
the number of fringes recorded at each temperature 
decreases as the critical temperature is approached, whereas the 
error made in counting the fringes is practically constant. Hence, 
in an effort to reduce the relative error on the fringe count for the 
data near the critical region, which are usually more significant, 
samples with longer light paths (5~and 10~mm) are used in that 
region. On the other hand, as data are taken farther and farther from 
the
critical point, the number of fringes at each datum increases greatly for 
the same light path ($\simeq 300$ fringes at $T_{\rm C}-T\simeq 0.1$~K with the 
10~mm cell), rendering 
the fringe counting operation more  uncomfortable and prone to 
 mistakes in counting. In the temperature region farther from \Tc\
the 2- and 1-mm light path sample cells are employed.

\subsection{Thermal Control}
Thermal control of required stability and uniformity is achieved in
two 
stages combined in a thermostat, whose detailed features are 
described in reference~\cite{nfthesis}.
The external stage is regulated within $\pm$ 0.1~K, or better, at a
temperature, $T_{\rm ext}$, such that $T_{\rm C}-T_{\rm ext}<$ 1 K,
by circulating thermostated water in a water jacket system. The second, inner,
stage is composed of a copper cylinder, to which heat is applied with 
heating foils, and of a cell holder, also a copper cylinder fitting 
snugly within the former. 
The inner cylinder and the cell holder are in turn placed inside
the external stage. After the second heating stage, the temperature
can be controlled within about $5\times 10^{-4}$ K. The temperature is
measured by a quartz thermometer and by a set of thermistors embedded
at various points in the copper cylinder. Both the quartz thermometer
and the thermistors were calibrated using a triple point cell to
reproduce the triple point of pure water. Tests were performed to 
ensure that adequate temperature uniformity and stability was 
achievable with this thermostat. By measurements done under different 
heating foil configurations, we estimated that thermal gradients 
along the height of the sample were less than $2\times 10^{-2}$~K/m,
which signifies a temperature difference of less than $6\times
10^{-5}$~K 
along the region of the meniscus, assuming it extends for a few mm in 
height. In terms of thermal stability, we monitored the temperature 
of the heater block, as measured by thermistors, for a time span of 
up to two weeks. In that time, the temperature of the experiment room 
increased by about 1.3$^{\circ}$C, while the temperature measured by 
the thermistors remained stable to within less than $4\times 
10^{-4}$~K.

\section{Results}
\subsection{The critical exponent $\beta$}
The collected coexistence curve data for the mixture {\it 
n}-heptane+nitrobenzene span 5 decades in terms of the 
reduced temperature $t$, from $t\simeq 4\times 10^{-7}$ to $t\simeq
4\times 10^{-2}$.
Data were taken so as to span the 
accessible range of temperatures of the mixture, which is around 
$12.5^{\circ}{\rm C}$ and is bounded by the 
difference between the critical consolute temperature and the
freezing temperature of {\it n}-heptane~\cite{merck}. The data are 
reported in the online supplemental file accompanying this article.

\begin{figure}\centering
\includegraphics[scale=0.45,angle=90]{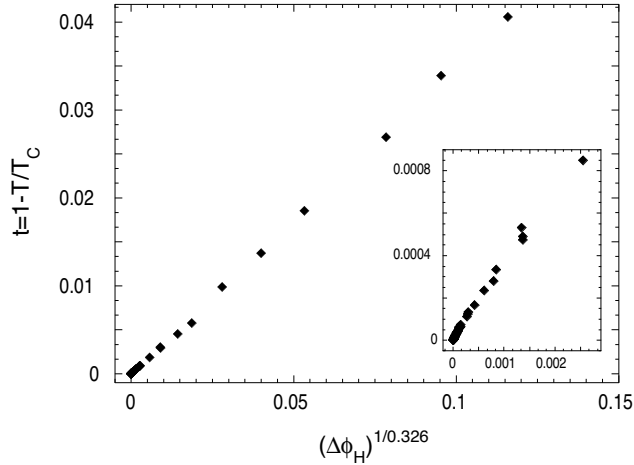}
\caption{The coexistence curve data for the binary
mixture {\it n}-heptane+nitrobenzene expressed as
$(\Delta\phi_{H})^{1/0.326}$ as a
function of the reduced temperature, where $\Delta\phi_{H}$ is the
{\it n}-heptane volume fraction.
The inset is an enlargement of the critical region.}\label{coexdataBL}
\end{figure}

The coexistence curve  data are shown in Fig.~\ref{coexdataBL}, where
they are expressed as $(\Delta\phi_{H})^{1/0.326}$ {\it vs} $t$.
 In this graph,
we have plotted $(\Delta\phi_{H})^{1/0.326}$ as a function of the reduced
temperature $t$, instead of the absolute temperature $T$, in order to
be able to report on the same graph data from several experimental
sequences with slightly different
critical temperatures.
We discuss below how the different 
critical temperatures arise.

We anticipated in the introduction that in the vicinity of 
the critical temperature, 
the coexistence curve is supposed to be described by the simple
scaling law:
\begin{equation}
\Delta\phi=B_0 t^{\beta}\label{simplescalingBL}
\end{equation}
where $t=1-T/T_{\rm C}$ and $\beta$ is the critical exponent.
The ``vicinity'' of \Tc, or $t=0$---the so-called asymptotic 
region---has generally been found to be the region with 
$t<10^{-2}$ for binary liquid mixtures~\cite{kumar83}. 
Translating this to the particular case of {\it 
n}-heptane+nitrobenzene, the region where the simple scaling law
should be 
valid extends to about 3~degrees below \Tc. It should be emphasized, 
however, that the asymptotic region is not known {\it a priori}, 
nor do we have a theoretical estimate of it. Moreover, in certain 
circumstances experiments on binary mixtures have suggested that the 
actual asymptotic region extends only to $t\simeq 10^{-3}$ from the 
critical temperature~\cite{greer76, jacobs96, aizpiri90}.
In view of such findings,  in the analysis of the 
present data a narrower asymptotic region $0<t<10^{-3}$
was chosen to find the critical temperature and the order parameter
critical 
exponent $\beta$ from the measured data.
Beyond the asymptotic region, correction-to-scaling terms become 
important:
\begin{equation}
\Delta\phi=B_0 t^{\beta}\left 
(1+B_1 t^{\Delta}+B_2 t^{2\Delta}+\cdots\right
)\label{corrtoscalingBL}
\end{equation}
where $\Delta$ is the correction-to-scaling critical exponent.

Because it is physically inaccessible, 
the critical temperature \Tc\
must inferred from the measured coexistence curve data. A first
estimate 
of \Tc\ is gathered by plotting the raw data as 
$(\Delta\phi)^{1/\beta}$ versus $T$ in the (expected) asymptotic
temperature 
range, where such a plot is linear. A linear fit to the data
intercepts the 
horizontal axis at \Tc. 
This value of the critical temperature is then taken 
as an initial value for a nonlinear least square fit of 
~(\ref{simplescalingBL}) to the coexistence curve data in the 
asymptotic region. In the fit,
$\beta$ and $B_{0}$ are used as free parameters.
(\Tc\ is only allowed to vary within the reasonable range suggested 
by a careful examination of the coexistence curve near 
the critical point.)
The best values of the parameters found by this nonlinear least square
fit are given in Table~\ref{simplescalingfitBL}.
\begin{table}
\begin{center}
    \begin{tabular}{ccccc} \hline\hline
	Fit & Region & $B_{0}$ & $\beta$ & $\chi^{2}$\\\hline
	A & $t{\mbox{\lower 1mm\hbox{$\stackrel{\textstyle <}{\sim}$}}} 
	10^{-3}$ & $1.34\pm 0.01$ & (0.326) & 
	$9.8\times 10^{-4}$\\
	B & $t{\mbox{\lower 1mm\hbox{$\stackrel{\textstyle <}{\sim}$}}} 
	10^{-3}$ & $1.91\pm 0.04$ & $0.367\pm 0.002$ & $1.3\times 
	10^{-4}$\\ \hline\hline
    \end{tabular}
    \caption{Parameters values for a nonlinear least square best fit 
    of $\Delta\phi=B_{0}t^{\beta}$ to the volume fraction data of 
    Fig.~\ref{coexdataBL}. Quantities in 
    brackets were held fixed during the
fit.}\label{simplescalingfitBL}
\end{center}    
\end{table}
As is apparent from the table, the fit obtained with $\beta$ set at 
the theoretical value of $\beta=0.326$ (fit A) does not seem to represent the 
data as well as the one where $\beta$ is unconstrained. 
Instead, the value of 
the apparent exponent, which produces the best fit is $\beta=0.367\pm 0.006$ 
(fit B).

\begin{figure}\centering
\includegraphics[scale=0.55]{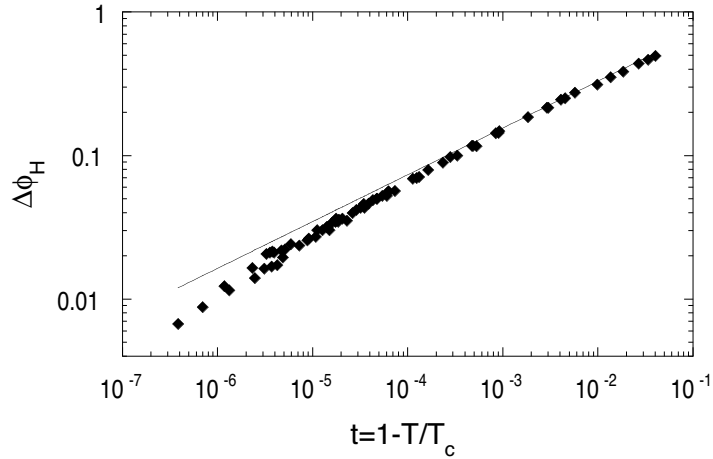}
\caption{Log-log plot of the order parameter $\Delta\phi_{H}$ versus 
the reduced temperature $t$. The slope of the data as $t$ tends to 
zero corresponds to the apparent exponent $\beta$. A line with slope 0.326 is 
also drawn for comparison.}\label{loglogBL}
\end{figure}

Once an estimate of \Tc\ has been obtained, a 
log-log plot of the data in the full temperature range 
is useful to see if any correction-to-scaling terms would be 
needed to interpret the data, and whether they should be positive or
negative. 
Also, the slope of the data in the asymptotic 
region in the log-log plot corresponds to the critical exponent.
A log-log plot for the data of Fig.~\ref{coexdataBL} is reported in 
Fig.~\ref{loglogBL}.
The slightly decreasing slope at higher values of $t$ indicates that 
some correction to the simple scaling law are needed and that it will 
have to have a negative coefficient. This can be seen by performing a 
nonlinear least square fit of the scaling law~(\ref{corrtoscalingBL}) 
to the coexistence curve data of Fig.~\ref{coexdataBL} 
in the whole range of temperatures 
studied, by using the best values found in the asymptotic range for
the 
critical temperature \Tc\ and for the critical exponent $\beta$
and 
leaving the coefficient $B_{1}$ free, while the 
correction-to-scaling exponent $\Delta$ is held fixed at its 
theoretical value of 0.54~\cite{chen82, zinn96}. The line through the 
data in Fig.~\ref{coexdataBL} corresponds to this fit and the 
parameters determined by the fit 
are reported in Table~\ref{corrtoscalingfitBL}, fit~C. 
\begin{table}
\begin{center}
    \begin{tabular}{cccccc} \hline\hline
	Fit & Region & $\beta$ & $B_{0}$ & $B_{1}$ & $\chi^{2}$\\\hline
	C & $t{\mbox{\lower 1mm\hbox{$\stackrel{\textstyle <}{\sim}$}}} 
	0.04$ & (0.367) & $1.905\pm 0.007$ & $-0.96\pm 0.03$ & $4.8\times 
	10^{-4}$\\
	D & $t{\mbox{\lower 1mm\hbox{$\stackrel{\textstyle <}{\sim}$}}} 
	0.04$ & (0.326) & $1.39\pm 0.01$ & $0.11\pm 0.07$ & $1.84\times 
	10^{-3}$ \\
	E & $t{\mbox{\lower 1mm\hbox{$\stackrel{\textstyle <}{\sim}$}}} 
	0.04$ & 0.361$\pm$0.002 & 1.82$\pm$0.03 & $-0.81\pm0.06$ & 
	$4.3\times 10^{-4}$\\ \hline\hline
    \end{tabular}
    \caption{Parameters values for a nonlinear least square best fit of 
    $\Delta\phi=B_0 t^{\beta}\left 
    (1+B_1 t^{\Delta}+B_2 t^{2\Delta}\right )$, with $t=(T-T_{\rm 
    C})/T$, to the volume fraction data of Fig.~\ref{coexdataBL}. Quantities
    in brackets were held fixed at the shown value during 
    the fit.}\label{corrtoscalingfitBL}
\end{center}    
\end{table}

It is evident from the best fit parameters of fits C and D that 
the scaling law with correction terms works better on the data when 
$\beta$ is the value found by the simple scaling law used in the 
asymptotic region (C) than when the theoretical value of $\beta$ is 
imposed during the fit (D). When the exponent is again treated as a 
free parameter in the fit, with one correction term (fit E), the 
value of $\beta$ agrees within error with the one found previously 
via the simple scaling law.

\begin{figure}\centering
\includegraphics[scale=0.55]{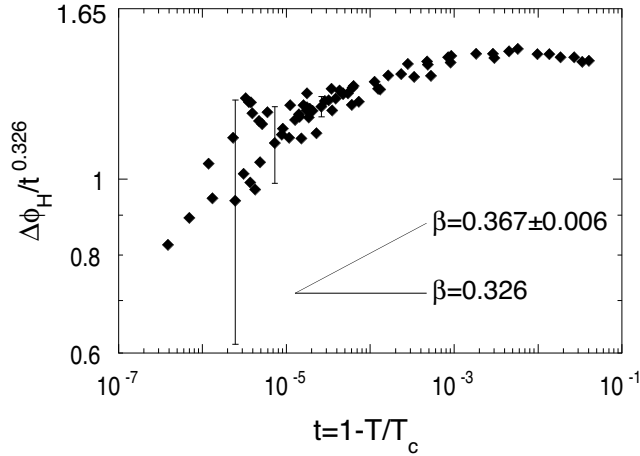}
\caption{Sensitive log-log plot, $\Delta\phi_{H}$ versus $t$,
of the coexistence data on {\it n}-heptane+nitrobenzene. 
The value used for the critical exponent is $\beta=0.326$. 
A sensitivity scale is also drawn in the graph to indicate the slope 
the data would preferentially take in the asymptotic region, 
were they plotted with the value 
of the exponent corresponding to the indicated
slope.}\label{sensitiveplotBL}
\end{figure}

A different way to perform the analysis of the data and arguably the 
ultimate verification of the importance of correction terms and of 
the goodness of the estimate of the measured $\beta$ is 
achieved by plotting the data in a manner that separates the 
contribution of the corrections terms from the rest of the terms in 
the relation~(\ref{corrtoscalingBL}).
If no correction terms are needed to fit the data and the value of 
the critical exponent is ``correct'', a log-log plot of 
$\Delta\phi_{H}/t^{\beta}$ {\it vs} $t$ 
would distribute the data along a horizontal line (we call this a 
{\it sensitive} plot). 
Departures from a zero-slope line would 
then indicate either an ``incorrect'' value of $\beta$ or the need of 
correction terms to fit the data, depending on the temperature range. 
Fig.~\ref{sensitiveplotBL}
is a sensitive plot of the coexistence data collected during the 
experiment with {\it n}-heptane+nitrobenzene, where the theoretical 
value of $\beta=0.326$ was used.
It is evident once again that a  value of $\beta$ larger than the 
theoretical value is necessary to flatten the slope of the data in 
the asymptotic region.
This type of graph can be used to perform a cross check on the values
of 
the critical temperature \Tc\ and $\beta$ produced by a 
nonlinear least square fit to the raw coexistence data of 
Fig.~\ref{coexdataBL}. Starting from some good guesses for 
\Tc\ and $\beta$, one can then vary each of them individually 
step-by-step until the data of the sensitive plot lies on a 
horizontal line. The critical temperature will only affect the data 
very close to $t=0$, while changes in $\beta$ will change the 
slope of the data on a wider range of $t$. 
The values of the critical temperature and 
the critical exponent produced by the experiment will then be the
values 
that make, within experimental error, a zero-slope plot, in the 
asymptotic region. Moreover, if after this stepwise
analysis 
the data appear distributed along different slopes at different
 ranges of $t$, this would be an indication that correction 
terms to the simple scaling law should be included in the fit.

The modern theory of critical phenomena being as 
widely accepted and successful as it is, it seemed 
worthwhile to try some fits to the data by using the theoretical
value 
of $\beta$ and adding {\it two} correction terms in the scaling 
law~(\ref{corrtoscalingBL}). It has been suggested that different 
exponents for the second correction term can be tried when analyzing 
coexistence curve data~\cite{fisherprivate, kumar83, greer86}.
Further fits 
to our coexistence data were then tried with the second correction 
terms being: $B_{2}t^{2\Delta}$, with $\Delta=0.54$, 
$B'_{2}t^{1-\alpha}$, where $\alpha=0.11$ is the specific 
heat critical exponent, and $B''_{2}t^{2\beta}$ with 
 $\beta=0.326$. The rationale behind keeping the 
choice of the second term open is based on the experimental
difficulty 
to distinguish between a correction exponent that is slightly 
larger ($2\Delta$) or smaller ($1-\alpha$ or $2\beta$) than one. 
The results of these fits are reported in 
Table~\ref{2ndcorrtoscalingfitBL} and they appear to produce worse 
results than the experimentally measured apparent value of
$\beta=0.367\pm0.006$.
\begin{table}
\begin{center}
    \begin{tabular}{ccccccc} \hline\hline
	Fit & Region & $\beta$ & $B_{0}$ & $B_{1}$ & $B_{2}$, $B'_{2}$, 
	$B''_{2}$& $\chi^{2}$\\ \hline
	G & $t{\mbox{\lower 1mm\hbox{$\stackrel{\textstyle <}{\sim}$}}} 
	0.04$ & (0.326) & $1.33\pm 0.03$ & $1.5\pm 0.9$ & $-7\pm 3$ & 
	$1.20\times 10^{-3}$\\ 
	H & $t{\mbox{\lower 1mm\hbox{$\stackrel{\textstyle <}{\sim}$}}} 
	0.04$ & (0.326) & $1.32\pm 0.03$ & $2\pm 1$ & $-6\pm 3$ & 
	$1.11\times 10^{-3}$\\ 
	I & $t{\mbox{\lower 1mm\hbox{$\stackrel{\textstyle <}{\sim}$}}} 
	0.04$ & (0.326) & $1.30\pm 0.05$ & $7\pm 3$ & $-9\pm 3$ & 
	$9.6\times 10^{-4}$\\ \hline\hline
    \end{tabular}
    \caption{Parameters values for a nonlinear least square best fit 
    of $\Delta\phi=B_0 t^{\beta}\left 
    (1+B_1 t^{\Delta}+B_2 t^{2\Delta}\right )$~(fit~G), 
     $\Delta\phi=B_0 t^{\beta}\left 
    (1+B_1 t^{\Delta}+B'_2 t^{1-\alpha}\right )$~(fit~H), and 
     $\Delta\phi=B_0 t^{\beta}\left 
    (1+B_1 t^{\Delta}+B''_2 t^{2\beta}\right )$~(fit~I) 
    to the volume fraction data of Fig.~\ref{coexdataBL}. Quantities
in 
    brackets were held fixed during the
fit.}\label{2ndcorrtoscalingfitBL}
\end{center}    
\end{table}

\subsection{The critical temperature \Tc}
Within each set of order parameter data the temperature was measured 
using the same thermistor, but different thermistors or other parts 
of the thermal control electronics were used in different sets. 
Moreover, as mentioned below in section~\ref{samples}, different 
samples were used to collect the body of data for this experiment.

Systematic experimental
studies~\cite{jacobs83, jacobs84} on the 
effect of water and acetone impurities in different binary liquid 
samples (methanol+cyclohexane) show 
that a percentage volume of water of about 0.1 in the mixture would
alter the 
critical temperature by almost +4~K, while 0.5\% of acetone gives 
a increase of about 2~K.
No regular variation of the critical temperature with the 
different samples was found and the different absolute values measured 
are all within a fraction of a percent of one another. While it is  
reasonable to think that both water and acetone impurities are
present 
in the samples (the glass manifolds used to fill the sample cells
are cleaned with both distilled 
water and acetone before making the samples), it is presumed that 
the amounts of those impurities do not 
differ much from one sample to the other. 
For these reasons, the actual critical consolute temperature 
of {\it 
n}-heptane+nitrobenzene varied somewhat: while for each data set it 
can be determined to within less than 0.5~mK, its absolute value from 
our measurements can 
only be given as $T_{\rm C}=(291.80\pm 0.02)\,{\rm K}$ 
(18.65~$^{\circ}{\rm C}$).

\section{Discussion}
Due to the marked departure of our measured value of the critical 
exponent $\beta$ from measurements on other systems as well as from 
the accepted theoretical value, we have carried out a 
careful analysis of sources of systematic errors, which 
may influence this type of measurements.
We believe that the influence of other sources of error on the 
accuracy of our data was minimized by
building an {\it ad hoc} improved thermostat and by being extremely
careful in the conduction of the experiments and in the preparation
of the samples.

\subsection{The effect of Earth's gravitational field}
\begin{figure}\centering
\includegraphics[scale=0.5]{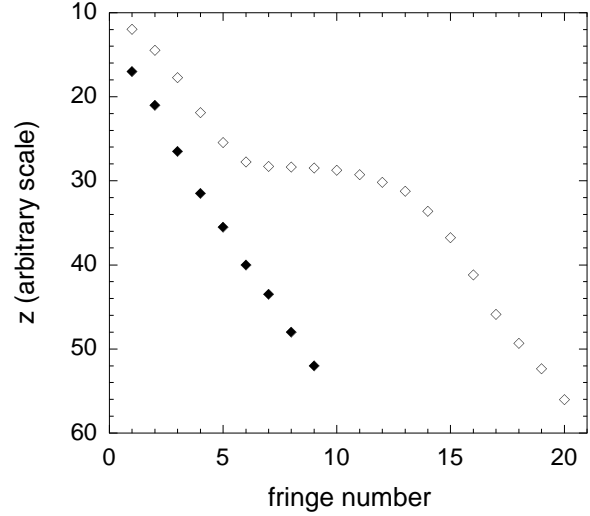}
\caption{Black diamonds: 
measurement of the refractive index profile at a temperature 
$T>T_{\rm C}$ after the sample was shaken to homogenize the 
phase. White diamonds: analogous measurement taken at a temperature 
$T<T_{\rm C}$ about 50 hours after the sample was cooled from
approximately $8\times 10^{-4}$ K above to
approximately $8\times 10^{-4}$ below \Tc.
}\label{nabovebelow}
\end{figure}

\begin{figure}\centering
\includegraphics[scale=0.45]{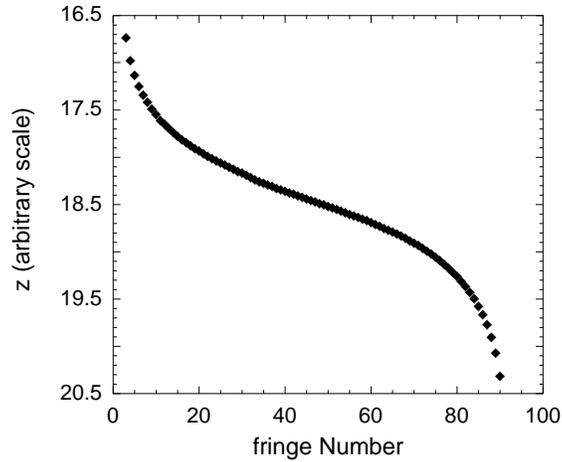}
\caption{Measurement of the refractive index profile at a temperature 
$T>T_{\rm C}$ after the sample was heated from about 1 mK below to about 1
mK above 
 critical, without shaking it.}\label{naboveTcsigmoid}
\end{figure}

Under the influence of Earth's gravitational field the refractive 
index vertical profile of a mixture below its consolute critical 
temperature is in principle distorted from a step-function shape like 
the one shown in 
Fig.~\ref{nprofile}a (references \cite{fannin74, greer75, giglio75, 
dickinson75} address this issue in binary liquid mixtures in the 
one-phase region).
It is conceivable that the gravity-modified profile 
would cause the appearance of diffraction fringes even when the 
system is at a temperature $T<T_{\rm C}$, if the distortion from a 
pure step-like profile were appreciable. 
Although, no ``unwanted'' fringes at temperatures below critical
were ever detected in any of the experimental data sets, 
to study this point in more depth, the image plane method was
employed because 
of its 
ability to map the index of refraction entire profile~(see 
section~\ref{optical}). By 
counting the fringes as a function of their position on the film 
(the latter being related to the height in the sample cell), the 
profile can be plotted.
Profile measurements were taken at a temperature above $T_{\rm C}$,
after 
the sample was shaken and therefore in a situation where a 
homogeneous index of refraction is expected, 
and below $T_{\rm C}$ roughly 50 hours after the temperature 
was lowered from above $T_{\rm C}$ (Fig.~\ref{nabovebelow}).
To ensure that this method is sensitive enough to 
reproduce a sigmoidal profile if one is present, a measurement was 
made of the profile after the temperature was raised from below to 
above $T_{\rm C}$, but without shaking the sample to homogenize the
phase. 
In this situation a profile like Fig.~\ref{nprofile}b is expected,
and 
is quite clearly measured by this method as is apparent from 
Fig.~\ref{naboveTcsigmoid}. 
From the measured profiles above and below the critical point, there 
seems to be no evidence of a dramatic deviation of the index of 
refraction from a homogeneous behaviour, both in the one-phase and 
the two-phase regions.

Gravitational effect are supposed to be more evident the larger the 
density difference between the two species of the
mixture~\cite{fannin74}. The 
densities of {\it n}-heptane and nitrobenzene ($\rho_{\rm H}=0.6837$
and 
$\rho_{\rm N}=1.2037$, at 20$^{\circ}$C referred to the 
density of water at 4$^{\circ}$C),
while not as closely matched as those of other
compounds~\cite{greer76}, 
are however less mismatched than at least one other mixture 
that has 
revealed no influence of gravitational effect on analogous 
measurements~\cite{jacobs96}.

\subsection{Equilibration}\label{equilibration}
Closely tied to the effect of the terrestrial gravitational field, 
a very important issue to address
is the equilibration time of
a binary mixture. As a consequence of the divergences occurring at
the critical point, the equilibration time also diverges as the set 
temperature $T$ tends to $T_{\rm C}$. 
Hence, the need to wait for an adequately long time after each 
temperature ($<$\Tc) is set before heating the system through \Tc\ to
obtain a 
datum. (And consequently the need for a very stable, as well as 
accurate thermal control system.) Several other authors have put 
considerable effort in the study of this phenomenon 
\cite{greer75, giglio75, dickinson75, fannin74, equilstudies}. 
From these studies, it appears that the most 
influential effect on equilibration is the sedimentation flux, which 
diverges as $(T-T_{C})^{-0.6}$. Only once in the searched literature, 
however, did we find reference to the effect of this divergence on 
the two-phase region of a binary mixture, in a study by D. Beysens 
(within \cite{equilstudies}). An order of magnitude calculation along the same 
lines suggests that in the asymptotic temperature range of our system 
the relaxation time due to sedimentation can reach several thousand 
hours (not unlike Beysens' findings), 
corroborating serious doubts on the feasibility of carring out 
meaningful critical phenomena studies of binary liquid mixtures on 
Earth. Clearly, this effect must have played a role in the 
experiments reporting results that are more compatible with the 
renormalization group theory predictions.

We employed the image plane method (section~\ref{optical}) 
to estimate the equilibration time of the 
system, after a temperature change. This was done by monitoring 
the interference fringes at the image plane of the focussing lens 
with time as the film was
moved in the camera. For our purposes, equilibrium was 
reached when the fringes  appeared horizontal on the film. 
In the most significative measurement performed, the temperature was 
set at a value $T_{\rm i}$ slightly above $T_{\rm C}$ and 
held there for about two hours, it was 
then decreased to a temperature $T_{\rm f}$ below $T_{\rm C}$, such
that 
$T_{\rm i}-T_{\rm f}{\mbox{\lower 1mm\hbox{$\stackrel{\textstyle 
<}{\sim}$}}}5\times 10^{-4}$~K. The system was kept at this
temperature and 
the fringes recorded on film for several days.
While the chart recorder track of the 
thermistors output showed that $T_{\rm f}$ was reached about
20~minutes after it 
was set, the fringes on the film do not appear to flatten to a 
horizontal slope until about 50~hours later.

The coexistence curve data close to $T_{\rm C}$ were all taken after
an equilibration time of about 50 hours, while for the rest of the data 
the equilibrium time allowance was between 10~and 20~hours. 

In  principle, data taken before equilibrium is reached would cause one 
to measure a smaller number of fringes than the ``true'' number at a 
given temperature, hence a smaller value of the order parameter 
$\Delta\phi$. This would, in turn, suggest that waiting a longer time 
would yield a more curved coexistence curve near $T_{C}$ and hence a 
``more correct'' value for $\beta$.

\subsection{Wetting}
There is some evidence that in binary mixtures one of the phases 
wets the walls of the cell containing the sample, as well as the 
other phase, sometimes surrounding the latter
completely~\cite{moldover80, 
degennes85}. 
The effect of this 
phenomenon in the experimental conditions of this work is that the 
effective thickness of the light path the laser traverses inside the 
cell is different from the nominal thickness given by the
manufacturer. 
Since the index of refraction, and hence the order parameter, is 
measured by the number of fringes detected at each temperature
divided 
by the light path length, if the latter is made uncertain by the 
presence of a wetting film, the consequent refractive index 
measurements will be affected and rendered unreliable.

By comparing refractive index 
measurements taken with cells of different thicknesses, the 
effect of a possible wetting layer can be monitored. Assuming the
wetting 
layer that forms has the same thickness regardless of the nominal
light 
path of the cell, if the data taken with different cells overlap
within experimental error, we can neglect 
the influence of the wetting layer.
The coexistence curve measurements were performed with different cell 
thicknesses to be able to map the whole range of $t$ available for
{\it n}-heptane+nitrobenzene with equal ease and accuracy. Those
measurements 
as reported in Fig.~\ref{loglogBL} and Fig.~\ref{sensitiveplotBL}
indicate
that the data indeed all seem to follow the same pattern.
Within the 
accuracy of the measurements the data overlap, indicating that if a 
wetting film is present, it is, however, unmeasurable in these 
experiments.

It is also worth emphasizing that there are indications from other 
observations~\cite{jacobs96, jacobs83, jacobs84} and theoretical 
predictions~\cite{cahn77}
that wetting behaviour occurs usually at temperatures several degrees 
below the critical temperature, a region of the coexistence curve 
with which one is less concerned when determining the critical 
exponent~$\beta$.

\subsection{Choice of \op}\label{mixingvolume}
As described above, we use index of refraction measurements to
extract the
volume fraction information to determine the coexistence curve of our
mixture.
In doing this three assumptions are made. The first is that the index 
of refraction does not present any singular behaviour at the critical
point.
There are theoretical studies~\cite{hartley74, sengers80, pepin88, 
beysens89} 
and experimental 
observations dealing with this issue, both showing 
that any anomaly in the refractive index at the critical point 
is below~100~ppm, less than the resolution of these experiments. In 
the case of \hn, measurements were made in earlier studies 
to ascertain the goodness of 
the proportionality relationship between the mixture volume fraction 
and the order parameter~\cite{sclavo}. 
The proportionality factor was found to be 
constant within about 0.1\% in the temperature range of the two-phase 
region of \hn. As such proportionality factor depends on the 
refractive index of the two phases of the mixture (see, for example, 
the reference within~\cite{jacobs76}), we feel safe in the assumption 
that in our system, too, any refractive index anomalies are likely 
not influential.

The second assumption made is that of zero mixing volume. In other 
words, when the two individual species, H and N, are joined 
together to form the mixture it assumed that their volumes, $V_{{\rm 
H}}$ and $V_{{\rm N}}$, simply add, 
while in general it is to be expected that 
$V_{{\rm mixture}}=V_{{\rm H}}+V_{{\rm 
N}}\pm|V_{{\rm E}}|$, where $V_{{\rm E}}$ is called the excess volume 
and it can be positive or negative. Experimental studies on two 
systems where $V_{{\rm E}}\neq 0$ have shown that this had no 
consequences on the measured critical exponent
$\beta$~\cite{jacobs86, reeder76}. 

Thirdly, it is assumed that the Lorentz-Lorenz relation remains 
valid for binary mixtures to the same extent as it valid for pure 
fluids, for which it holds within about 1\%, at least in the 
neighbourhood of the critical point. According to an 
experiment aimed at determining the validity of this 
assumption~\cite{jacobs86, houessou85, gastand90}, 
the Lorentz-Lorenz relation is verified 
within 0.5\%, when the volume loss upon mixing is considered.

We have 
assumed that the above assumptions are valid for the results obtained 
with this system, although no measurements were carried out on it 
to verify such assumptions.

\section{Summary and conclusions}
Our measurement of the coexistence curve power law critical exponent
 for the mixture \hn\ as the temperature
$T\rightarrow T_{c}$ yields an apparent value of $\beta=0.367\pm 
0.006$, consistently higher than the accepted
theoretical value of $\beta_{\rm theory}\simeq 0.326\pm 0.002$ 
\cite{guida98}. In an
effort to find a reason for this discrepancy in some experimental
flaw, the known potential sources of systematic error were carefully
analyzed. In this process, the image plane optical technique 
was employed to measure the profile of the index of
refraction of the binary liquid sample in a way that we had not
tried before. The results are interesting in that the shape of the
refractive index vertical gradient can be mapped directly by this
technique. The  relatively new use of this optical tool, as well as
employment of an improved thermal control apparatus for the reported
measurements have helped rule out that surface wetting by one of the
phases, the influence of other effects such as impurities in the
samples, and the exact definition of
the order parameter
could be responsible for the discrepancy between the measured and the
theoretically predicted exponent $\beta$. A quick literature survey 
revealed one other recent report of a similarly high value for 
$\beta$ from an experiment \cite{venkatesu05}. However, in that work, 
there appears no attempt  to 
find a reasonable explanation for the result.

Perhaps, two effects remain 
as candidate culprits for the high value of $\beta$ we measured. 
On the one hand is the equilibration time issue. 
It is somewhat unclear (from this work and 
previous literature \cite{greer75, giglio75, dickinson75, fannin74, 
equilstudies}) whether and to what 
extent the influence of the gravitational field on the equilibration 
time of this mixture would affect the final outcome of our experiments. 
Estimates of 
equilibration times for a given system may vary by an order of 
magnitude (see, for example, remarks in \cite{giglio75} about 
observations obtained in \cite{greer75}). Even in the least 
consevative case, those 
estimates would indicate that our equilibrium time allowances were 
too short and hence that some of our measurements, closer to the 
critical point, were performed before equilibrium was reached. 
Should that be indeed the case, while it would leave open a potential 
explanation for the discrepancy between our measured value of $\beta$ 
and its widely accepted value, we would still be unable 
to explain why the fringes observed in the image plane measurements, 
aimed at establishing when equilibrium was reached,
do become horizontal after a time supposedly shorter than the 
mixture's equilibration time (section \ref{equilibration}).

On the other hand, it is conceivable that the asymptotic 
temperature range of this system be much closer to the 
critical point than expected for binary mixtures (or even for pure 
fluids). Our measurements in this case would suggest that 
the asymptotic region for this binary mixture would have to 
extend only to $t<10^{-5}$, a considerably narrower interval than the more 
commonly observed $t<10^{-3}$. 
\vspace{5mm}

{\bf Acknowledgments}
\vspace{5mm}

We wish to thank Arman Bonakdarpour for his help in the conduction of
parts 
of the experiment. Many thanks also to the department technicians, 
Doug Wong and Joe O'Connor for their great help in making our glass 
manifolds, and to Tim Dainard and Bill Goldring for the distillation
of the mixture samples. N.~F. wishes to thank Professor Ian
Affleck for helpful discussions.
This work was supported in part by National Sciences and Engineering
Research Council (NSERC) of Canada grants to D.~A.~B.


\begin{thebibliography}{99}
\bibitem{ordparamnote} To answer the question of which kind of concentration 
is best suited for the order parameter, the established trend is to use an 
order parameter the quantity that renders the coexistence curve more symmetric, 
in order to be able to compare experimental results with the lattice gas model 
predictions\cite{kumar83}. Experience suggests that the volume 
fraction usually yields more symmetric coexistence curves. Moreover, 
the quantity that is usually measured in our binary liquid 
experiments is the index of refraction difference between 
the two phases, which turns out to be proportional to the volume concentration.
\end{thebibliography}
\end{document}